\documentclass[12pt]{article}
\usepackage{amssymb,amsmath,epsfig}
\allowdisplaybreaks

\begin{document}

\title{\bf Stable Charged Gravastar Model in ${f}(\mathfrak{R},\textbf{T}^{2})$ Gravity with Conformal Motion}
\author{M. Sharif \thanks {msharif.math@pu.edu.pk} and Saba Naz
\thanks{sabanaz1.math@gmail.com}\\
Department of Mathematics, University of the Punjab,\\
Quaid-e-Azam Campus, Lahore-54590, Pakistan.}

\date{}
\maketitle
\begin{abstract}
This paper investigates the influence of charge on physical features
of gravastars in the framework of energy-momentum squared gravity. A
gravastar is an alternate model to a black hole that comprises of
three distinct regions, namely the intermediate shell, inner and
outer sectors. Different values of the barotropic equation of state
parameter provide the mathematical formulation for these regions. We
construct the structure of a gravastar admitting conformal motion
for a specific model of energy-momentum squared gravity. The field
equations are formulated for a spherically symmetric spacetime with
charged perfect matter distribution. For the smooth matching of
external and internal spacetimes, we use Israel matching criteria. Various
physical attributes of gravastars such as the equation of state
parameter, proper length, entropy and energy are investigated (in
the presence of charge) versus thickness of the shell. The charge in
the inner core of gravastars preserves the state of equilibrium by
counterbalancing the inward gravitational force. It is concluded
that the non-singular solutions of charged gravastar are physically
viable in the background of energy-momentum squared gravity.
\end{abstract}

\section{Introduction}

The study of the incredibly huge cosmos gives insight into its
mysterious nature. Recent cosmic observations such as supernova type
1a, cosmic microwave background radiations as well as large-scale
structures indicate that the cosmos is expanding at a rapid rate
\cite{sc}. The unknown strange force responsible for this expansion
is dubbed as dark energy. Although the general theory of relativity
(GR) provides a platform for the discussion of cosmic issues, it
fails to explain the accelerated expansion of the universe due to
the well-known problems: fine-tuning and cosmic coincidence.
Moreover, the occurrence of singularities in GR is a crucial problem
as they are predicted in the regime of high-energy where GR is not
applicable due to quantum effects. Based on these issues, different
modifications of GR have been proposed to study the singularity
problem as well as to investigate the attributes of mysterious dark
energy. A natural extension of GR, known as $f(\mathfrak{R})$
gravity, is obtained by including the associated function of
curvature invariant $(\mathfrak{R})$ in the geometric sector of the
Einstein-Hilbert action. Different cosmological and astrophysical
problems have been successfully addressed in this theory \cite{3a}.

Harko et al. \cite{4a} introduced $f(\mathfrak{R},T)$ gravity as a
generalization of $f(\mathfrak{R})$ theory by including the trace of
energy-momentum tensor (EMT) $T$. Haghani et al. \cite{5a}
incorporated the contraction of Ricci scalar and EMT
$({\mathfrak{R}}_{\mu\nu} {T}^{\mu\nu})$ in $f(\mathfrak{R},T)$
gravity to analyze the effect of non-linear correction terms
appearing due to the non-minimal coupling between curvature and
matter. These theories disobey the conservation law which proves the
existence of an additional force acting on particles. Thus, under
the influence of this force, the motion of particles is
non-geodesic. These theories have captivated the attention of many
researchers as the minimal/non-minimal interaction of matter and
curvature is thought to be a good fit for comprehending the
the connection between rapid expansion and dark cosmic components.
Moreover, these couplings efficiently describe the rotation curves
of galaxies and distinct cosmic ages of the cosmos.

The beginning of the cosmos presents an interesting problem for
cosmologists. In this regard, different researchers have tried to
formulate adequate cosmic models explaining the evolution of the
universe since the advent of time. According to the big-bang theory,
matter and energy of the cosmos were initially concentrated at a
single point (known as a singularity) having limitless density and
temperature with no surface area or volume. The universe came into
existence due to the explosion of superheated ultra-dense matter
within the singularity and has been expanding since then. Big bounce
theory is an alternative approach describing the evolution of the
cosmos as a series of big bounces (expanding and contracting over and
over again) having no beginning or end state. In this regard,
energy-momentum squared gravity (EMSG) \cite{7a} has been proposed
which favors the big bounce theory. It modifies the Einstein-Hilbert
action by including the analytic function of the form
${T}_{\mu\nu}{T}^{\mu\nu}=\textbf{T}^{2}$. In the perspective of
this theory, the cosmos with maximum energy density as well as a
minimum-scale factor in the early epoch bounces to a state of
expansion \cite{8a}. Consequently, this theory resolves the big-bang
singularity in non-quantum prescription without affecting cosmic
evolution.

The quadratic components of matter source in EMSG field equations
help in resolving fascinating cosmological and astrophysical issues.
In the context of the standard cosmological model, the cosmological
constant does not play an important role in the early times and becomes
important only after the matter-dominated era. Roshan and Shojai \cite{8a}
found that in this theory, the repulsive nature of
the cosmological constant plays a crucial role at early times in
resolving the singularity. Moreover, it has been shown that this theory possesses a true
sequence of cosmological eras. Board and Barrow \cite{5m} found isotropically
expanding cosmos for matter theories of higher-order. Nari and Roshan \cite{6m}
determined the existence of pressureless compact stars in EMSG and
obtained stellar models that are less compact in comparison to their
GR counterparts. Akarsu et al. \cite{7m} discussed the modification
of the $\Lambda$CDM model. They investigated the non-minimal
interactions for relativistic compact objects and dark matter in the
context of EMSG and have obtained suitable constraints on the model.
Moraes and Sahoo \cite{8m} studied the geometry of non-exotic
wormholes. Bahamonde et al. \cite{9m} presented the expansion of
cosmos using different coupling models and found that these models
are a good fit to explain the current behavior of cosmos. Sharif and
Gul \cite{10m} used Noether symmetries in this theory and studied
the impact of different physical parameters on viable cosmological
structures. They also inspected the viability as well as stability of
compact celestial bodies and concluded that the correction terms of the
gravity decrease the collapse.

Different astrophysical phenomena, such as the development and
evolution of stellar structures have sparked the interest of
different researchers. Among all the cosmic entities, stars are the
core components of galaxies, which are arranged systematically in a
cosmic web. The outward pressure in a star vanishes when it exhausts
its fuel. Consequently, the star undergoes gravitational collapse
which leads to the formation of compact objects. One of the stellar
remnants is a black hole which is a completely collapsed object with
a singularity covered by an event horizon. In order to overcome
singularity and event horizon problems, Mazur and Motolla \cite{13m}
proposed a cold compact model (referred to as gravitationally vacuum
star or gravastar) as an alternative to the black hole. The most crucial
feature of this hypothetical object is its singularity-free nature
in contrast to the black hole. It comprises of three regions: the
interior and exterior regions are separated by an intermediate
thin-shell. The singularity is avoided by considering the de Sitter
(dS) spacetime in the innermost region with a surrounding layer of
cold baryonic fluid that separates the inner and outer regions,
whereas the exterior vacuum is described by the Schwarzschild
metric. Moreover, each sector is characterized by a particular
equation of state (EoS).

There is currently no observational evidence in support of
gravastar, however, a few indirect evidences in literature can be
used to predict gravastar's occurrence and future detection. Sakai
et al. \cite{d1} presented a mechanism to detect gravastar through
the exploration of gravastar shadows. Gravitational lensing is
another possible approach for detecting gravastars, as hypothesized
by Kubo and Sakai \cite{d2}, who claimed that microlensing effects
of maximal luminosity do not occur in black holes. However, these
effects can be observed in gravastar having the same mass as that of
black hole. The detection of GW150914 through interferometric LIGO
detectors \cite{d3}-\cite{d4}, hinted towards the presence of
ringdown signal generated by objects without an event horizon. A recent
analysis of the image acquired by the First M87 Event Horizon
Telescope (EHT) suggests a shadow that could belong to a gravastar
\cite{d5}.

Visser and Wiltshire \cite{14m} inspected the stability of
gravastars against radial perturbations and found that the use of
feasible EoS leads to stability of gravastar in GR. Carter
\cite{15m} extended this work by analyzing suitable constraints for
the stability of non-singular exact solutions of gravastars. Catteon et
al. \cite{15mm} investigated the usual structure of gravastars and
found that the configuration possesses anisotropy in the absence of
shell. Bilic et al. \cite{15mmm} constructed solutions of gravatars
by using inner Born-Infeld phantom spacetime in place of dS geometry
and formulated solutions representing compact objects at the core of
galaxies. Horvat and Iliji\'{c} \cite{16m} examined the stability of
the gravastar by applying the speed of sound criterion on the shell
and obtained surface compactness bounds. Ghosh et al. \cite{cd}
showed that the extension of 4-dimensional gravastar to higher
dimensions is not possible. Researchers have also investigated the
existence and important physical features of gavastars in the
context of modified theories of gravity \cite{29aa}-\cite{65aa}.

Conformal symmetry describes the inherent symmetry of Killing
vectors. Mathematically, it is given in the form
\begin{equation}
\nonumber L_{\varsigma}g_{m n}=\Phi(r)g_{m n},
\end{equation}
where the operator $L_{\varsigma}$ describes the Lie derivative,
$\Phi$ is the conformal factor and $\varsigma$ denotes a
four-vector. The conformal Killing vectors reduce highly nonlinear
field equations to linear ordinary differential equations. Herrera
and Leon \cite{17} used exact solutions corresponding to spherically
symmetric spacetime and found a parametric group of conformal
motions. Banerjee et al. \cite{7ddd} investigated various attributes
of a braneworld gravastar that allowed conformal motion.

Electromagnetism plays a crucial role in the study of the evolution
as well as stability of collapsing objects. A star requires an
immense amount of charge to counterbalance the inward pull of
gravity and sustain its equilibrium state. Lobo and Arellano
\cite{cg1} studied gravastar solutions incorporating nonlinear
electrodynamics and examined significant structural attributes.
Horvat et al. \cite{cg2} investigated charged gravastar and
calculated surface redshift, sound speed as well as the EoS
parameter for the model under consideration. Turimov et al.
\cite{cg3} presented a brief discussion on slowly rotating
gravastars filled with highly magnetized perfect matter. Usmani et
al. \cite{cg4} used conformal motion to compute gravastar models
with charged interior and Reissner-Nordstr\"{o}m as the exterior. They
concluded that the charged interior region acts as an
electromagnetic mass model which generates gravitational mass thus
contributing to the stability of the gravastar structure. Rahaman et
al. \cite{cg5} studied numerous viable features for gravastar
structures in 3-dimensional spacetime with the contribution of
charge. Sharif and Javed \cite{7dddd} investigated the stable
configuration of gravastars by considering quintessence as well as
regular black hole geometries in the interior and observed linear
profiles of physical variables versus the thickness of the shell.

In the background of $f(\mathfrak{R}, T)$ gravity, Sharif and Waseem
\cite{cgc7} investigated charged gravastar structure using conformal
motion and found that the inner sector behaves as an electromagnetic
mass model in the presence of charge which helps in the formation of
singularity-free structure. Bhatti et al. \cite{cg7} explored the
stability of charged gravastar in the modified gravity. Bhar and Rej
\cite{cg77} studied gravastar admitting conformal motion to analyze
the contribution of charge to the stability region for the discussed
model. Bhatti and his collaborators \cite{28a} also studied
gravastar structure in the context of ${f(\mathcal{G})}$ gravity,
where $\mathcal{G}$ is Gauss-Bonnet invariant. They investigated
different attributes of gravastar with and without electromagnetic field.
Various attributes of gravastar solutions are also analyzed
through the gravitational decoupling technique in \cite{28aa}.
Sharif and Naz \cite{kk} studied gravastar structure with Kuchowicz
metric in ${f}(\mathfrak{R},\textbf{T}^{2})$ gravity.
Recently, Sharif and Saeed \cite{ii} studied charge-free
gravastar accepting conformal motion in background of
$f(\mathfrak{R},\textbf{T}^{2})$ gravity and discussed the behavior
of various physical attributes. Motivated by numerous works
presenting effects of charge, we are interested to analyze the
influence of charge on the gravastar model in
$f(\mathfrak{R},\textbf{T}^2)$ gravity.

In the present work, we examine the gravastar configuration
admitting conformal motion for charged spherical static system in
$f(\mathfrak{R},\textbf{T}^2)$ gravity. The paper is organized
according to the following outline. Section \textbf{2} describes the
framework of $f(\mathfrak{R},\textbf{T}^2)$ gravity and presents
field equations with conformal motion. In section \textbf{3}, we
investigate three distinct regions of gravastar with different EoS.
Section \textbf{4} includes the discussion of physical attributes of
gravastar, i.e., EoS parameter, proper length, entropy and energy of
the shell. The last section concludes the major findings of the
work.

\section{Basic Formalism of $f(\mathfrak{R},\textbf{T}^{2})$ Theory}

In this section, we develop the field equations of
$f(\mathfrak{R},\textbf{T}^{2})$ theory and conformal equations for
charged isotropic matter distribution. The action of this theory is
described as \cite{7a}
\begin{equation}\label{1}
\mathcal{I}=\frac{1}{2\kappa^2}\int
d^4x\left[f\left(\mathfrak{R},{\textbf{T}}^{2}
\right)+\mathcal{L}_{e}+\mathcal{L}_{m}\right]{\sqrt{-g}},
\end{equation}
where $g$ indicates the determinant of the metric tensor,
$\mathcal{L}_{m}$ symbolizes the matter Lagrangian and $\kappa^2=1$
is a coupling constant. Furthermore,
$\mathcal{L}_{e}=\frac{-1}{16\pi}\mathcal{F}^{\mu\nu}\mathcal{F}_{\mu\nu}$
where $\mathcal{F}_{\mu\nu}=\varphi_{\nu,\mu}-\varphi_{\mu,\nu}$
describes the electromagnetic field tensor and $\varphi_{\mu}$
represents the four potential. The respective field equations turn
out to be
\begin{equation}\label{2}
\mathfrak{R}_{\mu\nu}f_{\mathfrak{R}}-\frac{1}{2}g_{\mu\nu}f+g_{\mu\nu}\Box
f_{\mathfrak{R}}-\nabla _{\mu}\nabla_{\nu}f_{\mathfrak{R}}
=T_{\mu\nu}-\Theta_{\mu\nu}f_{\textbf{T}^{2}}+\mathrm{E}_{\mu\nu},
\end{equation}
where $f\equiv f(\mathfrak{R},\textbf{T}^{2})$,$~ \Box={\nabla_{\mu}
\nabla^{\mu}}, ~f_{\mathbf{T}^{2}}=\frac{\partial f}{\partial
\textbf{T}^2}, f_{\mathfrak{R}}=\frac{\partial f}{\partial
\mathfrak{R}}$,
\begin{eqnarray}\label{3}
\Theta_{\mu\nu} =-4\frac{\partial^{2}\mathcal{L}_{m}}{\partial
g^{\mu\nu}
\partial g^{\theta\eta}}T^{\theta\eta}-2\mathcal{L}_{m}\left(T_{\mu\nu}-\frac{1}{2}g
_{\mu \nu }T\right)-TT_{\mu \nu}+2T_{\mu}^{\theta}T_{\nu\theta},
\end{eqnarray}
and the electromagnetized energy-momentum tensor
$\mathrm{E}_{\mu\nu}$ is provided by
\begin{equation}\label{}
\mathrm{E}_{\mu\nu}=
\frac{1}{4\pi}\left(\mathcal{F}^{\theta}_{\mu}\mathcal{F}_{\nu\theta}-\frac{1}{4}g_{\mu\nu}
\mathcal{F}^{\theta\eta}\mathcal{F}_{\theta\eta}\right).
\end{equation}
Non-conserved EMT is generated by  $f(\mathfrak{R},\textbf{T}^{2})$
gravity, implying the existence of an additional force that
describes non-geodesic motion of particles as follows
\begin{equation}\label{4}
\nabla^{\mu}T_{\mu\nu}=\frac{1}{2} \left[-f_{
\mathbf{T}^{2}}g_{\mu\nu}{\nabla^{\mu}}{\mathbf{T}^{2}}+2{\nabla^{\mu}}(f_{
\mathbf{T}^{2}}{\Theta_{\mu\nu}})+\nabla^{\mu}\mathrm{E}_{\mu\nu}\right].
\end{equation}
In gravitational physics, the EMT describes the energy and matter
distribution, where the non-vanishing components provide physical
variables that describe various dynamical properties.

We consider isotropic matter configuration
\begin{equation}\label{5}
T_{\mu\nu}=v_{\mu} v_{\nu}(\mathcal{P}+\varrho)-\mathcal{P}
g_{\mu\nu},
\end{equation}
where $ v_{\mu}$, $\varrho$ and $\mathcal{P}$ define the
four-velocity, energy density and pressure of the fluid,
respectively. For matter configuration, there are two choices of
$\mathcal{L}_{m}$, i.e., either $\mathcal{L}_{m}=\mathcal{P}$, or
$\mathcal{L}_{m}=-\varrho$. For theories possessing minimal
interaction, matter lagrangian choice does not influence the
distribution \cite{pp}. Here, we consider
$\mathcal{L}_{m}=\mathcal{P}$ and utilizing Eqs.(\ref{3}) and
(\ref{5}), we have
\begin{eqnarray}\nonumber
\Theta_{\mu\nu} = -\left(\varrho^2+3\mathcal{P}^2+4
\mathcal{P}\varrho\right)v_{\mu}v _{\nu}.
\end{eqnarray}
The field equations (\ref{2}) are complex as well as non-linear in
the framework of multivariate functions and their differentials.
Therefore, we choose a particular function that demonstrates
\emph{minimal/non-minimal} interactions among geometry and matter
components. The field equations corresponding to the non-minimal
model turn out to be very complex making the solution of these
equations more challenging. Thus, we choose a minimal model for our
analysis \cite{7a}
\begin{equation}\label{6}
f(\mathfrak{R},\textbf{T}^{2})= \mathfrak{R}+\beta \textbf{T}^{2},
\end{equation}
where $\mathbf{T}^{2}=\varrho^{2}+3\mathcal{P}^{2}$ and $\beta$ is
the model parameter. It describes the current accelerated expansion
and cosmic evolution of the cosmos. Incorporating $\textbf{T}^{2}$
into this modified gravity results in a wider form of GR than
$f(\mathfrak{R})$ and $f(\mathfrak{R},T)$ theories. The functional
form (\ref{6}) explains the $\Lambda$CDM model and has frequently
been used to investigate different aspects of self-gravitating
objects. It solves a variety of cosmological problems \cite{ab} and
corresponds to three major epochs of the universe, i.e.,
dS-dominated, radiation-dominated and matter-dominated era. Plugging
this model in Eq.(\ref{2}), we have
\begin{equation}\label{9}
\mathbb{G}_{\mu\nu}= T_{\mu\nu}+\frac{1}{2}\beta g_{\mu\nu}
{\textbf{T}}^{2}-\beta f_{
{\textbf{T}}^{2}}{\Theta_{\mu\nu}}+\mathrm{E}_{\mu\nu},
\end{equation}
where $ \mathbb{G}_{\mu\nu}$ is the Einstein tensor. We choose
static spherically symmetric geometry to examine the interior region
of the charged gravastar as
\begin{equation}\label{9a}
ds^{2}=e^{\chi(r)}dt^{2}-e^{\lambda(r)}dr^{2}-r^{2}(d\theta^{2}+\sin^{2}\theta
d\phi^{2}).
\end{equation}
The associated field equations are
\begin{eqnarray}\label{10}
&&\left[-\frac{e^{-\lambda}}{r^{2}}+ \frac{1}{r^{2}}+
\frac{\lambda^{'}e^{-\lambda}}{r}\right]=\varrho + \frac{3}{2}\beta
\varrho^{2} + \frac{9}{2} \beta \mathcal{P}^{2} +4 \beta \varrho
\mathcal{P}+\frac{{q}^{2}}{{8\pi{r}^{4}}},
\\\label{11}
&&\left[e^{-\lambda}\left(\frac{1}{r^{2}}+\frac{\chi^{'}}{r}\right)
- \frac{1}{r^{2}}\right]=\mathcal{P} - \frac{\beta}{2}\varrho^{2}
-\frac{3}{2} \beta
\mathcal{P}^{2}-\frac{{q}^{2}}{{8\pi{r}^{4}}},\\\nonumber && \left[
e^{-\lambda}\left(\frac{\chi^{'2}}{4}+\frac{\chi^{'}}
{2r}-\frac{\lambda^{'}}{2r}+\frac{\chi^{''}}{2}
-\frac{\chi^{'}\lambda^{'}}{4}\right) \right]=\mathcal{P}
-\frac{\beta}{2}\varrho^{2}-\frac{3}{2}\beta
\mathcal{P}^{2}+\frac{{q}^{2}}{{8\pi{r}^{4}}},\\\label{12}
\end{eqnarray}
where prime corresponds to the radial derivative.

The conformal killing vectors are obtained by using conformal symmetry
which is defined as
\begin{equation}\nonumber
L_{\varsigma}g_{m
n}=g_{jn}\varsigma^{j}_{;m}+g_{mj}\varsigma^{j}_{;n}=\Phi(r)g_{m n}.
\end{equation}
Solving above, we have
\begin{equation}\nonumber
\varsigma^{1}\chi^{'}=\Phi ,\quad \varsigma^{1}=\Phi\frac{ r}{2}
,\quad \varsigma^{1}\lambda^{'}+2\varsigma^{1}_{1}=\Phi, \quad
e^{\lambda}=\left(\frac{b}{\Phi}\right)^{2}, \quad
e^{\chi}=a^{2}r^{2} ,
\end{equation}
where $a$ and $b$ are constants of integration. Inserting the obtained
results in Eqs.(\ref{10})-(\ref{12}), we have
\begin{eqnarray}\label{13}
\left[\frac{1}{r^2}\left(1-\frac{2\Phi
\Phi^{'}r}{b^2}-\frac{\Phi^2}{b^2}\right)\right]&=&\varrho +
\frac{3}{2}\beta \varrho^{2} + \frac{9}{2} \beta \mathcal{P}^{2} +4
\beta \varrho \mathcal{P}+\frac{{q}^{2}}{{8\pi{r}^{4}}}
,\\\label{14}\left[\frac{1}{r^2}\left(\frac{3\Phi^2}{b^2}-1\right)\right]&=&\mathcal{P}
-\frac{\beta}{2}\varrho^{2}-\frac{3}{2}\beta
\mathcal{P}^{2}-\frac{{q}^{2}}{{8\pi{r}^{4}}},\\\label{15}
\left[\frac{1}{b^2 r^2}\left(\Phi^2+2\Phi
\Phi^{'}r\right)\right]&=&\mathcal{P}
-\frac{\beta}{2}\varrho^{2}-\frac{3}{2}\beta
\mathcal{P}^{2}+\frac{{q}^{2}}{{8\pi{r}^{4}}},
\end{eqnarray}
where $q$ denotes the charge of sphere specified by
\begin{equation}\label{}
q(r)=\int_{0}^{r}4 \pi r^{2}e^{\lambda/2}\sigma(r)dr,\quad
E=\frac{q}{4\pi r^{2}}.
\end{equation}
Here $\sigma$ and $E$  represent the charge density of surface and
electric field intensity, respectively. The solution of the
non-conserved equation (\ref{4}) provides
\begin{equation}\label{17}
\frac{d\mathcal{P}}{dr}+\frac{\chi'}{2}(\varrho+\mathcal{P})-\frac{qq'}{4\pi
r^{4}}+\aleph^\star=0,
\end{equation}
where $\aleph^\star$ shows the additional impact of energy-momentum
squared gravity as follows
\begin{equation}\nonumber
\aleph^\star=\beta\frac{\chi'}{2}\left(3\mathcal{P}^2+\varrho^2+4
\mathcal{P}\varrho\right)+\beta(\varrho\varrho'+3\mathcal{P}\mathcal{P}').
\end{equation}

To reduce the complexity of respective field equations, we employ
standard barotropic EoS expressed by
\begin{equation}\label{100}
\mathcal{P}=\omega\varrho,
\end{equation}
where $\omega$ corresponds to EoS parameter. The corresponding expressions of
energy density, pressure and charge are
\begin{eqnarray}\nonumber
\varrho&=&\frac{1}{2 \sqrt{\pi } b^2 \beta  r^4 (9 \omega ^2+8
\omega +3)}\Bigg[\Bigg\{b^2 r^4 (4 \pi  b^2 r^4-\beta (9 \omega ^2+8
\omega+3) \\\label{19}&\times&(b^2 (q^2-8 \pi r^2)-16 \pi r^3 \Phi
\Phi'+8 \pi  r^2 \Phi^2))\Bigg\}^{\frac{1}{2}}-2 \sqrt{\pi }
b^2 r^4\Bigg],\\\nonumber \mathcal{P}&=&\frac{1}{b^2 \beta  r^2 (3
\omega ^2+5)} \Bigg[ \Bigg\{b^2 r^2 \omega ^2 (b^2 (3 \beta \omega
^2+5 \beta +2 r^2 \omega ^2)-2 \beta  r
\\\label{19aa}&\times&(3 \omega ^2+5)\Phi \Phi'-4 \beta (3 \omega ^2+5) \Phi^2)
\Bigg\}^{\frac{1}{2}}\sqrt{2} +2 b^2 r^2 \omega^2\Bigg], \\\nonumber
q&=&\frac{\sqrt{8 \pi }}{(3 \omega ^2+5)\sqrt{{-b^2\beta }}}
\Bigg[-\Bigg\{r^2 (b^2 (\beta (9 \omega ^4+18 \omega ^2+5)+6
r^2(\omega^2
\\\nonumber&-&
1) \omega ^2)+8 \beta r (3 \omega ^2+5)
\Phi \Phi'+\beta  \left(-27 \omega ^4-42 \omega ^2+5\right)
\\\nonumber &\times&\Phi^2)+3\sqrt{2} (\omega ^2-1)\times\Bigg(b^2 r^2 \omega ^2 (b^2(\beta (3 \omega ^2+5)
+2 r^2 \omega ^2)\\\label{nan}&-&2 \beta  r (3 \omega ^2+5)
\Phi\Phi'
-4 \beta  (3 \omega ^2+5) \Phi^2)\Bigg)^{\frac{1}{2}}\Bigg\}
\Bigg]^{\frac{1}{2}}.
\end{eqnarray}
For $\beta=0$, all of these equations are reduced to those of GR
accepting conformal Killing vectors equations of motion \cite{cg4}.

\section{Solutions of Gravastar Model}

The structure of charged gravastars involves the study of three
different regions: (1) the interior geometry (2) the exterior
geometry (3) thin-shell. The inner region is enclosed by
intermediate shell comprising of stiff fluid, whereas
the outermost sector is fully vacuum described by
Reissner-Nordstr\"{o}m metric. The entire structure
of gravastar including impact of charge can be presented as\\\\
(A) Interior region $(\mathcal{S}_{1})$ $(0 \leq
r<r_{1}=\mathcal{R})$ \quad$ \Longrightarrow\quad
\varrho=-\mathcal{P} $,\\\\
(B)Intermediate shell $(\mathcal{S}_{2})$ $(r_{1}\leq r\leq
r_{2}=\mathcal{R}+\epsilon)$
 \quad $\Longrightarrow \quad\varrho=\mathcal{P} $,\\\\
(C) Exterior region $(\mathcal{S}_{3})$ $(r_{2}<r)$ \quad
$\Longrightarrow\quad
\varrho=0=\mathcal{P}$.\\\\
The radii of internal and external regions are indicated with
 $r_{1}$ and $r_{2}$,
respectively which measures the thickness of the shell such as
$r_{2}-r_{1}=\epsilon$.

\subsection{Interior Geometry}

In the core, we consider $\omega=-1$ to demonstrate the repulsive
behavior of pressure. This value is regarded to be the best fit for
studying the effects of dark energy. Consequently, the connection
between respective energy density and pressure becomes
\begin{eqnarray}\nonumber
\varrho+\mathcal{P}&=&\frac{1}{4b^2\beta r^4}\Bigg[r^2\Bigg\{{b^2
r^2(b^2\Xi-8\beta r\Phi\Phi'-16\beta\Phi
^2)}\Bigg\}^{\frac{1}{2}}\\\label{n1}&+&\Bigg\{{b^2 r^6 (b^2
\Xi-24 \beta r\Phi\Phi')}\Bigg\}^{\frac{1}{2}}\Bigg]=0,
\end{eqnarray}
where $\Xi=4 \beta +r^2$. Thus, in the interior region,
Eq.(\ref{n1}) provides a conformal factor of the form $\Phi=
\Phi_{0} r$ with $\Phi_{0} $ as a dimensionless integration
constant. The corresponding physical parameters are provided as
follows
\begin{eqnarray}\label{20}
\varrho&=&\frac{\sqrt{b^4 \left(r^8+4 \beta  r^6\right)-24 b^2 \beta
{\Phi}_{0}^2 r^8}-b^2 r^4}{4 b^2 \beta  r^4}=-\mathcal{P},
\\e^{-\lambda}&=&\tilde{\Phi}_{0}^{2} r^{2}, \quad e^{\chi}=a^{2}r^{2},
\\ E^{2}&=&\frac{q^{2}}{16\pi^{2} r^{4}}=\frac{1}{4\pi
r^{2}}, \\\sigma&=&\frac{\tilde{\Phi}_{0}}{r\sqrt{4\pi}},
\end{eqnarray}
with $\tilde{\Phi}_{0}=\frac{\Phi_{0}}{b}$ having dimension
$\frac{1}{L}$. The gravitational mass of the inner core of charged
gravastar model is specified as
\begin{equation}\label{21}
M=\int_0^{r=\mathcal{R}}
4\pi(\varrho+2\beta\varrho^2+\frac{{q}^{2}}{{8\pi{r}^{4}}} )r^{2}dr.
\end{equation}

\subsection{The Intermediate Geometry}

The intermediate shell is sandwitched between two spacetimes, i.e.,
exterior and interior spacetimes. It comprises of stiff fluid that
obeys the EoS ($\mathcal{P}=\varrho$). The notion of stiff matter
configuration was developed by Zeldovich \cite{29}. This type of
matter distribution has been used by a group of researchers and have
obtained interesting results about cosmological and astrophysical
structures \cite{59aa}- \cite{63aa}. By employing the corresponding
EoS, we obtain matter variables of charged thin-shell as follows
\begin{eqnarray}\label{n10}
\varrho&=&\frac{\sqrt{b^2 r^6 (b^2 (20 \beta +r^2)-120 \beta r \Phi \Phi')}-b^2 r^4}{20 b^2 \beta r^4},\\\label{102}
\mathcal{P}&=&\frac{\sqrt{b^2 r^2 (b^2 (\Xi)-8 \beta r \Phi
\Phi'-16 \beta  \Phi^2)}+b^2 r^2}{4 b^2 \beta  r^2},\\
q&=&\frac{2 \sqrt{\pi } \sqrt{-r^2 (b^2+2 r \Phi \Phi '-2
\Phi^2)}}{\sqrt{-b^2}}, \\\nonumber
\varrho-\mathcal{P}&=&\frac{1}{20 b^2 \beta  r^4}\Bigg[ -5 r^2
\Bigg\{{b^2 r^2 (b^2 (\Xi)-8 \beta  r \Phi \Phi'-16 \beta
\Phi^2)}\Bigg\}^{\frac{1}{2}}\\\label{101}&+&\Bigg\{{{b^2 r^6 (b^2
(20 \beta +r^2)-120\beta r\Phi\Phi ')}}\Bigg\}^{\frac{1}{2}}-6 b^2
r^4\Bigg]=0.
\end{eqnarray}
\begin{figure}\center\epsfig{file=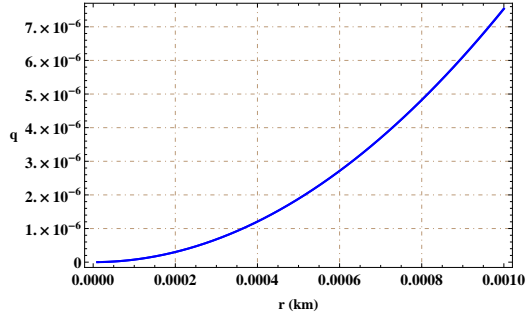,width=0.5\linewidth}
\caption{Graphical analysis of charge and thickness for $b=0.3$.}
\end{figure}
It is notable that the electric field is unaffected by the model
parameter and has the same form as in GR.  The charged shell's
thickness is incredibly thin and ranges between 0 and 1, i.e.,
$0<e^{-\lambda}\ll1$. Moreover, the respective equations to this
region are non-linear higher-order differential equations. Thus,
obtaining an exact solution to the field equations is difficult. We
have numerically solved the differential equation (\ref{101}) by
using the initial condition $\Phi(0) = 0.1$ and obtained an
interpolating function for the conformal factor corresponding to
$b=0.3$ and $\beta=0.1$. The plot of charge against the thickness of
shell has positive behavior as presented in Figure \textbf{1}. Thus,
the presence of charge provides an additional repulsive force that
resists gravitational collapse leading to the formation of a more
stable structure as compared to the uncharged model.

\subsection{The Exterior Geometry and Israel Formalism}

The Reissner-Nordstr\"{o}m metric corresponds to this vacuum region obey the
EoS $\varrho=\mathcal{P}=0$ and is expressed as follows
\begin{equation}\label{24}
ds^{2}=\curlyvee(r)dt^{2}-\curlyvee(r)^{-1}dr^{2}-r^{2}
(d\theta^{2}+\sin^2{\theta}d\phi^{2}),
\end{equation}
where $\curlyvee(r)=1-\frac{2 \mathcal{M}}{r}+\frac{\mathcal{Q}^{2}}
{r^{2}}$, $\mathcal{M}$ is the total gravitational mass of gravastar
and $\mathcal{Q}$ represents corresponding total charge. Its interesting
to evaluate the constraints that will allow us to match the internal
and external regions smoothly. In this context, Israel's
formalism plays key role in achieving the appropriate conditions of
smooth matching. The metric coefficients must be continuous at the
hypersurface $(\Sigma)$, i.e. at $r=\mathcal{R}$ but their
derivatives might be discontinuous. We use Lanczos equations to
determine the surface stress-energy tensor given as follows
\begin{equation}\label{25}
S_{\beta}^{\alpha}=(\delta_{\beta}^{\alpha}
\tau_{\gamma}^{\gamma}-\tau_{\beta}^{\alpha})\frac{1}{8\pi},
\end{equation}
where $\tau_{\alpha\beta}=K^{+}_{\alpha\beta}-K^{-}_{\alpha\beta}$.
It symbolizes the discontinuity in extrinsic curvature. The exterior
and interior regions are characterized by positive and negative
signs, respectively. Extrinsic curvature components at $(\Sigma)$
are represented as
\begin{equation}\label{26}
{K_{\alpha\beta}^{\pm}}= -n_{\zeta}^\pm \left[\frac{\partial^2
x^{\zeta}}{\partial \xi^{\alpha}
\xi^{\beta}}+\Gamma^{\zeta}_{\lambda\delta}\left(\frac{\partial
x^{\lambda}}{\partial\xi^{\alpha}}\right)\left(\frac{\partial
x^{\delta}}{\partial\xi^{\beta}}\right)\right],
\end{equation}
where $\xi^{\beta} $ indicates the coordinates of intrinsic shell
and $n_{\zeta}^{\pm}$ shows the unit normal at $\Sigma$. It is provided by
\begin{equation}\label{27}
n_{\zeta}^{\pm}=\pm\left|g^{\lambda\delta}\frac{\partial
\curlyvee}{\partial x^{\lambda}}\frac{\partial \curlyvee}{\partial
x^{\delta}}\right|^{-\frac{1}{2}}\frac{\partial \curlyvee}{\partial
x^{\zeta}}, \quad n_{\zeta} n^{\zeta}=1.
\end{equation}

For charged perfect fluid matter configuration, EMT takes the
follwing form $S_{\beta}^{\alpha}= \text{diag}(\Psi, -\Omega,
-\Omega)$, where $\Psi$ and $\Omega$ are surface energy density and
surface pressure provided by Lanczos equations as
\begin{eqnarray}\label{28}
\Psi&=&-\frac{1}{4\pi
\mathcal{R}}\left[\sqrt{\curlyvee}\right]_{-}^{+},\\\label{29}
\Omega&=&
-\frac{\varrho}{2}+\frac{1}{16\pi}\left[\frac{{\curlyvee^{'}}}
{\sqrt{\curlyvee}}\right]_{-}^{+}.
\end{eqnarray}
We obtain matter variables by inserting the metric functions of
interior and exterior spacetimes in the preceding equations which
leads to
\begin{eqnarray}\label{30}
\Psi&=&-\frac{1}{4 \pi \mathcal{R}}\left[\sqrt{1-\frac{2\mathcal{M}}
{\mathcal{R}}+\frac{\mathcal{Q}^{2}}{\mathcal{R}^{2}}}-\tilde{\Phi}_{0}\mathcal{R}\right],
\\\label{30}\Omega&=&\frac{1}{8\pi
\mathcal{R}}\left[\frac{\mathcal{R}-\mathcal{M}}
{\mathcal{R}\sqrt{1-\frac{2\mathcal{M}}{\mathcal{R}}+\frac{\mathcal{Q}^{2}}{\mathcal{R}^{2}}}}-
2\tilde{\Phi}_{0}\mathcal{R} \right].
\end{eqnarray}

Employing energy density at surface, the mass of the
intrinsic shell is obtained as
\begin{equation}\nonumber
\mathcal{M}_{shell}=4\pi
\mathcal{R}^{2}\Psi=\left[\tilde{\Phi}_{0}\mathcal{R}-\sqrt{1-\frac{2
\mathcal{M}}{\mathcal{R}}+\frac{\mathcal{Q}^{2}}{\mathcal{R}^{2}}}\right]\mathcal{R}.
\end{equation}
As a result, the total mass of charged gravastar in terms of
$\mathcal{M}_{shell}$ takes the form
\begin{equation}\label{32}
\mathcal{M}=\frac{1}{2\mathcal{R}}\left[\mathcal{R}^{2}-\tilde{\Phi}_{0}^{2}\mathcal{R}^{6}
-\mathcal{M}^{2}_{shell}{\mathcal{R}}^{2} +2
\tilde{\Phi}_{0}\mathcal{M}_{shell}\mathcal{R}^{4}+\mathcal{Q}^{2}
\right].
\end{equation}

\section{Some Physical Attributes of Charged Gravastars}

The physical characteristics of charged gravastar, such as the EoS
parameter, entropy, energy, and appropriate length, are discussed in
this section.

\subsection{The EoS Parameter}

It describes the connection between matter variables, such as
pressure and energy density of fluid configuration. Taking into
account the effective variables of matter contents, we get the EoS
parameter at $r=\mathcal{R}$ with following mathematical description
$\omega(\mathcal{R})=\Omega/\Psi $. Substituting the corresponding
values of matter variables, we obtain
\begin{equation}\label{33}
\omega(\mathcal{R})=\frac{\Omega}{\Psi}=\frac{\frac{1}{8\pi
\mathcal{R}}\left[\frac{\mathcal{R}-\mathcal{M}}
{\mathcal{R}\sqrt{1-\frac{2\mathcal{M}}{\mathcal{R}}+\frac{\mathcal{Q}^{2}}{\mathcal{R}^{2}}}}-
2\tilde{\Phi}_{0}\mathcal{R} \right]}{-\frac{1}{4 \pi \mathcal{R}}\left[\sqrt{1-\frac{2\mathcal{M}}
{\mathcal{R}}+\frac{\mathcal{Q}^{2}}{\mathcal{R}^{2}}}-\tilde{\Phi}_{0}\mathcal{R}\right]}.
\end{equation}
This equation includes different fractional components as well as
square root terms that makes the equation more complex. To
attain real EoS parameter, the basic requirement is either $\frac{2
\mathcal{M}}{\mathcal{R}}-\frac{\mathcal{Q}^{2}}{\mathcal{R}^{2}}<1$,
or $\frac{2 \mathcal{M}}{\mathcal{R}}<1$. The obtained relation
among $\mathcal{M}$, $\mathcal{R}$ and $\mathcal{Q}$ is
$\mathcal{Q}>\sqrt{\mathcal{R}(2\mathcal{M}-\mathcal{R})}$ and
$\mathcal{M}<\frac{\mathcal{R}}{2}$.
 It is observed that
positive pressure and energy density correspond to a positive value
of the EoS parameter and $\omega (\mathcal{R})\approx 1$ is acquired
for higher values of $\mathcal{R}$ \cite{19}.  These values of
$\mathcal{R}$ illustrate the perspective of cold compact objects like
gravastars and the absence of pressure makes it equivalent to a dust
shell.
\begin{figure}
\center \epsfig{file=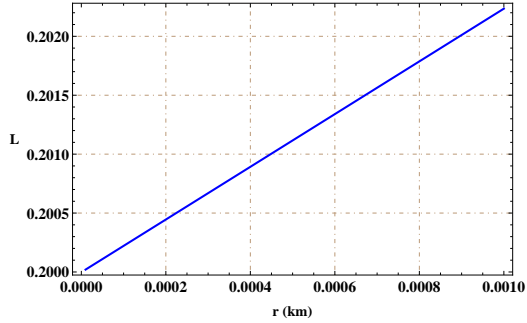,width=0.5\linewidth}
\caption{Graphical trend of the proper length and thickness for
$b=0.3$ and $\beta=0.1$.}
\end{figure}

\subsection{Proper Length}

The proper length of charged gravastar is measured from interior
region boundary  $r_{1}=\mathcal{R}$ to the thin-shell boundary
$r_{2}=\mathcal{R}+\epsilon$, where $\epsilon$ is incredibly small
thickness of charged shell. Consequently, we have
\begin{equation}\label{n6}
{L}=\int_{\mathcal{R}}^{\mathcal{R}+\epsilon}\sqrt{e^{\lambda}}dr.
\end{equation}
Differentiation of Eq.(\ref{n6}) provides
\begin{equation}\label{n7}
L'(r)=\frac{b}{\Phi}\Big|_{\mathcal{R}}^{\mathcal{R}+\epsilon}.
\end{equation}
In order to graphically analyze the behavior of length, we use the
numerical solution of $\Phi$ and solve Eq.(\ref{n7}) for the
initial condition $L(0)=0.2$. The length of the charged thin-shell
shows an increasing profile against the thickness of the shell as
shown in Figure \textbf{2}.

\subsection{Entropy of Shell}

Within the core of charged gravastar, entropy is a measure of
randomness or disorderness in the respective region. The entropy
density of the interior region is found to be zero in the literature
\cite{13m}. Inside the charged shell, the entropy is given by
\cite{33}
\begin{equation}\label{35}
S=4\pi\int_{\mathcal{R}}^{\mathcal{R}+\epsilon}\sqrt{e^{\lambda}}
r^{2} \verb"s"(r)dr,
\end{equation}
where entropy density is denoted by $\verb"s"(r)$ and its
mathematical description is as follows
\begin{equation}\label{36}
\verb"s"(r)=\frac{\Upsilon^{2}w_{\beta}^{2}{T}(r)}{4\pi{\hslash^2}}=
\Upsilon\left(\frac{w_{\beta}}{\hslash}\right)\sqrt{\frac{\mathcal{P}}{2\pi}},
\end{equation}
where $\Upsilon$ is a scalar constant, $T(r)$ is the temperature,
$\hslash$ is Planck's constant and $w_{\beta}$ is Boltzmann
constant. As changing the value of a constant simply scales the
graph, therefore the physical trend of the variables is not
affected. For convenience, we take $\hslash=w_{\beta}=1$.
Consequently, entropy turns out to be
\begin{equation}\label{37}
S=\int_{\mathcal{R}}^{\mathcal{R}+\epsilon}\frac{4\pi \Upsilon
r^2}{\sqrt{2\pi}}\sqrt{\mathcal{P} e^\lambda}dr.
\end{equation}
Substitution Eq.(\ref{102}) and  $e^\lambda=(\frac{b}{\Phi})^2$
leads to the following form
\begin{equation}
S=\int_{\mathcal{R}}^{\mathcal{R}+\epsilon}\frac{
b\sqrt{8\pi}\Upsilon}{\Phi}\Bigg[\frac{{\sqrt{b^2 r^2 (b^2 (\Xi)-8
\beta r \Phi \Phi '-16 \beta \Phi^2)}}}{4 b^2 \beta
r^2}+\frac{1}{4\beta}\Bigg]^{\frac{1}{2}}r^2 dr,\\\label{37a}
\end{equation}
whose differentiation provides
\begin{equation}
S'(r)=\frac{ b\sqrt{8\pi}\Upsilon}{\Phi}\Bigg[\frac{\sqrt{b^2 r^2
(b^2(\Xi)-8 \beta r \Phi \Phi '-16 \beta \Phi^2)}}{4 b^2 \beta
r^2}+\frac{1}{4\beta}\Bigg]^{\frac{1}{2}}r^2\Big|_{\mathcal{R}}^{\mathcal{R}+\epsilon}.\\\label{n3}
\end{equation}
The interpolating function of entropy is obtained by numerically
solving Eq.(\ref{n3}) along with the initial condition $S(0)= 0.2$.
The plot shows that the entropy of shell increases. Thus, charge and
the correction terms of $f(\mathfrak{R},\mathbf{T}^2)$ gravity play
a key role in increasing the randomness of the geometrical structure
(refer to Figure \textbf{3}).
\begin{figure}
\center \epsfig{file=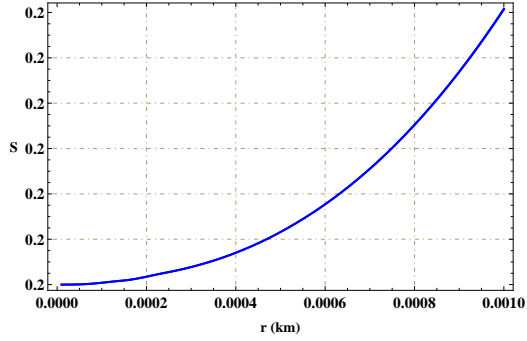,width=0.5\linewidth}\caption{Behavior of shell's entropy versus thickness of the shell for $b=0.3$ and
$\beta=0.1$.}
\end{figure}

\subsection{Shell Energy}

The charged gravastar's inner region has repulsive behavior and
meets the following EoS $\varrho=-\mathcal{P}$. In charged
gravastar, this repulsive character avoids singularity formation,
while the energy contained within thin-shell is illustrated as
\cite{32}
\begin{equation}\label{39}
\mathcal{E}=\frac{1}{2}\int_{\mathcal{R}}^{\mathcal{R}+\epsilon}
\left(\varrho+2\pi {E}^{2}\right) r^{2}dr,
\end{equation}
which simplifies to
\begin{eqnarray}\nonumber
\mathcal{E}&=&\int_{\mathcal{R}}^{\mathcal{R}+\epsilon}
\left(\frac{\sqrt{b^2 r^6 \left(b^2 \left(20 \beta +r^2\right)-120
\beta  r \Phi \Phi '\right)}}{40 b^2 \beta  r^2}-\frac{r^2}{40\beta
}\right.\\\label{40}&+&\left. \frac{40 r \Phi \Phi '}{40b^2}
-\frac{40 \Phi ^2}{40b^2}+\frac{1}{2}\right)dr.
\end{eqnarray}
Differentiation of the above equation leads to
\begin{equation}\label{n4}
\mathcal{E}'(r)=\frac{\sqrt{b^2 r^6 \left(b^2 \left(20 \beta
+r^2\right)-120 \beta  r \Phi \Phi '\right)}}{40 b^2 \beta
r^2}-\frac{r^2}{40\beta }+\frac{ r \Phi \Phi '}{b^2} -\frac{ \Phi
^2}{b^2}+\frac{1}{2}\Big|_{\mathcal{R}}^{\mathcal{R}+\epsilon}.
\end{equation}
The numerical solution of above equation is obtained by using
initial condition $\epsilon(0)=0.2$ which is plotted in Figure
\textbf{4}. It describes the increasing trend of the intrinsic shell
energy against $\epsilon$.
\begin{figure}\center\epsfig{file=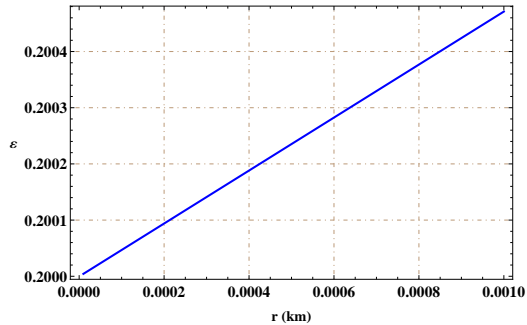,width=0.5\linewidth}
\caption{Graphical trend of shell's energy and thickness for
$b=0.3$ and $\beta=0.1$.}
\end{figure}

\section{Final Remarks}

In this paper, we have investigated the influence of charge on an
alternate model of a black hole, dubbed as gravastar in EMSG. For this
purpose, we have considered a setup admitting conformal motion
corresponding to a minimally coupled linear model, i.e.,
$f(\mathfrak{R},\textbf{T}^{2})$=$\mathfrak{R}+\beta
\textbf{T}^{2}$. We have adopted a linear EoS
$\varrho=\omega\mathcal{P}$ to inspect different regions of charged
gravastar. For the interior dS spacetime, we take $\omega=-1$ to
incorporate negative pressure which maintains the balance of forces
and avoids the formation of a singularity. The shell of small
thickness corresponds to the intermediate region filled with
baryonic matter and obeying $\omega=1$. We have represented the
vacuum exterior of the charged gravastar through
Reissner-Nordstr\"{o}m black hole satisfying
$\varrho=\mathcal{P}=0$. Israel junction conditions have provided
surface density and pressure which help to determine the mass of
charged thin-shell as well as the total gravitational mass. We have
graphically inspected the role of electromagnetic field for
different values of $\beta$. In the current scenario, physically
viable and stable geometry of gravastar is obtained for
$\frac{2\mathcal{M}}{\mathcal{R}}-\frac{\mathcal{Q}^{2}}{\mathcal{R}^{2}}<1$.
Figures \textbf{2}-\textbf{4} demonstrate that the proper length,
entropy and energy increases against the thickness of the shell in
the presence of charge.

We have found that the length, entropy and energy of the gravastar
structure increase in the presence of charge. Also, in other
modified theories of gravity, the graphical analysis of various
features has shown that the physical attributes of the shell follow
an increasing trend \cite{30aa}-\cite{tt}. Thus, our analysis
provides consistent results with other modified theories of gravity.
Moreover, as the charge provides a positive outward-directed force,
therefore, an additional repulsive force helps the gravastar from
collapsing into a singularity. Thus, the presence of charge
generates a more stable structure as compared to the uncharged
model. Our analysis follows the consistent increasing trend of
length, energy and entropy of charged analogs presented in GR as
well as other modified theories of gravity
\cite{7dddd,cgc7,cg77,28a}. Furthermore, we have found that under
the influence of ${f}(\mathfrak{R},\textbf{T}^{2})$ gravity, the
physical attributes of gravastar have greater values of length,
energy and entropy in comparison to work presented in GR and
${f(\mathfrak{R}, T)}$ gravity \cite{7dddd,cgc7}. Recently, Sharif
and Saeed \cite{ii} studied charge-free gravastar accepting
conformal motion in background of $f(\mathfrak{R},\textbf{T}^{2})$
gravity and concluded that proper length decreases while entropy as
well as energy increases abruptly. In contrast to the uncharged
case, in the presence of charge proper length, energy and entropy
increase with respect to thickness of the shell. We conclude that
$f(\mathfrak{R},\textbf{T}^{2})$ gravity satisfactorily discusses
charged gravastar model admitting conformal motion and provides more
stable structure in comparison to uncharged analog \cite{ii}. It is
noteworthy that our results reduce to GR for $\beta=0$
\cite{cg4}.\\\\
\textbf{Data availability:} No new data were generated or analyzed
in support of this research.

\end{document}